\documentclass[a4paper,10pt]{article} 
\input epsf.tex

\newcommand{\be}{\begin{equation}}
\newcommand{\ee}{\end{equation}}
\newcommand{\ba}{\begin{eqnarray}}
\newcommand{\ea}{\end{eqnarray}}
\newcommand{\ban}{\begin{eqnarray*}}
\newcommand{\ean}{\end{eqnarray*}}

\newcommand{\ket}[1]{\mbox{$ | #1 \rangle $}}
\newcommand{\bra}[1]{\mbox{$ \langle #1 | $}}

\newcommand{\si}{\sigma}
\newcommand{\demi}{\frac{1}{2}}

\newcommand{\one}{\leavevmode\hbox{\small1\normalsize\kern-.33em1}}

\begin{document}

\title{\Large \sc{Superluminal influences, hidden variables, and signaling}}
{\normalsize{\author{ Valerio Scarani\thanks{corresponding author.
Tel. +41 22 7026883; fax +41 22 7810980; e-mail:
valerio.scarani@physics.unige.ch}, Nicolas Gisin
\\Group of Applied Physics, University of Geneva\\
20, rue de l'Ecole-de-M\'edecine, CH-1211 Geneva, Switzerland\\
 }}}
\date{}
\maketitle

\begin{abstract}
We consider alternative models to quantum mechanics, that have
been proposed in the recent years in order to explain the EPR
correlations between two particles. These models allow in
principle local hidden variables produced at the source, and some
superluminal "hidden communication" (or "influences") to reproduce
the non-local correlations. Moving to the case of three particles,
we show that these alternative models lead to signaling when
"hidden communication" alone is considered as the origin of the
correlations.
\end{abstract}

\section{Introduction}

In the classical world, we are accustomed to two kinds of causes
for correlations: (i) A manufacturer produces white and black
balls; he sells them by pairs, and he always puts a white and a
black ball in each box. Here, the source of the perfect
anticorrelation is pre-established in the fabric. (ii) A referee
decides that the match is over, and sends a signal (whistle) to
all players, who stop running. The correlated behavior of the
players is the result of having received a signal. However,
neither of these models can explain the {\em
Einstein-Podolski-Rosen (EPR) correlations}, that are, the
correlations between two quantum entangled particles. In fact, on
the one hand, these correlations are established for space-like
separated events, ruling out the explanation through the reception
of a common signal \cite{asp,tittel,weihs}; on the other hand,
their explicit dependence on the meaningful parameters of the
experiments cannot be the consequence of sharing a common
information pre-established at the source --- this is Bell's
theorem \cite{bell,nobell}.

The physics community is accustomed since more than one century to
the counter-intuitive character of some predictions of quantum
mechanics (QM). Still, facing EPR correlations, some physicists
are not satisfied: as John Bell put it, "correlations cry out for
explanation", and not only for "description". That's why in the
last decades some models have been proposed that provide a
dynamical explanation to quantum correlations between two
particles \cite{bohm,Pearle,GRW,PSD,HPA,sua}. These models
incorporate Bell's theorem: the correlations are not entirely
pre-established at the source; but they admit the existence of
some form of communication between the particles, that we shall
call here {\em hidden communication}. This communication must
propagate at a speed $v_{hc}$ which is superluminal, to be
consistent with experiments performed in the configuration of
space-like separation. But one cannot define a superluminal
communication in a relativistic space-time \cite{reich}, without
allowing the causal loop or backward-in-time signaling. Therefore
the model must single out the frame in which the hidden
communication occurs.

Among these dynamical models, some are "alternative models", since
the explanation that is proposed leads also to predictions that
{\em differ} in some cases from those of QM, without contradicting
the experimental data that are available at the time of the
proposal. These models are worth of consideration: they lead to
design new experiments, that can discriminate between them and
standard QM. If the experiments falsifies QM, a door is opened for
new physics; but even if QM is confirmed, which is probable {\em a
priori}, the investigation of the alternative model will have shed
new light on the intimate connection between QM and relativity.

We mentioned the astonishing features of EPR correlations, but
there is another element that prevents these correlations to
become something too dramatic. In spite of their "non-local"
character, EPR correlations do not allow {\em signaling}, that is,
sending a message at an arbitrary high speed. More precisely,
these correlations alone cannot be used to send a signal at any
speed; and classical communication must be established between the
two observers if they want to check their correlations. Now, any
alternative model, in addition to reproduce QM for all the
experiments performed to date, is also supposed to fulfill the
no-signaling condition. A model that fails to comply with this
requirement would be in open contradiction with special
relativity, thus acquiring a very problematic status --- although
conclusive falsification can come only through experiment.

In this paper, we consider alternative models built on the idea of
superluminal hidden communication. We show that if correlations
are built only by this communication, these models imply
signaling. Thus, any alternative model must invoke both
superluminal hidden communication and local variables.

We begin with some general considerations about the no-signaling
condition, and the constraints that it imposes.

\section{From two to three particles}
\label{twothree}

Consider two quantum particles of arbitrary spin prepared in an
entangled state, then flying apart from one another, each to one
observer, Alice or Bob. Each observer makes a measurement. Let's
call $a_j$, $j=1,...,J$ the possible outcomes for Alice's
measurement; and $b_k$, $k=1,...,K$ the possible outcomes for
Bob's measurement. The joint probability for two outcomes is
written $P(a_j,b_k)$; the marginal distributions on Alice's and
Bob's side are $P(a_j)=\sum_kP(a_j,b_k)$, respectively
$P(b_k)=\sum_jP(a_j,b_k)$.

Now, Alice could decide to make another measurement, whose
outcomes are labelled $a_l'$, with $l=1,...,L$. The no-signaling
condition here means that $P'(b_k)=\sum_lP(a'_l,b_k)$ must be
equal to $P(b_k)$: on his own side, Bob cannot notice that Alice
has chosen another measurement. Of course, the symmetric condition
must yield on Alice's side: Alice's marginal distribution cannot
be modified by Bob's choosing another measurement.

Now, within the framework of QM, "another measurement" means to
measure another observable, or more generally to perform a
different POVM. But here we are willing to consider alternative
models to QM, that give different predictions for the EPR
correlations. In this case, "another measurement" may mean that
Alice modifies her parameters so that the correlations are no
longer those predicted by QM, that is $P(a_j,b_k)\neq
P^{QM}(a_j,b_k)$. Therefore, in the more general framework in
which we work, the no-signaling condition implies also that
$P(a_j)=P^{QM}(a_j)$, and $P(b_k)=P^{QM}(b_k)$.

\begin{figure}[t]
\begin{center}
\epsfxsize=9cm \epsfbox{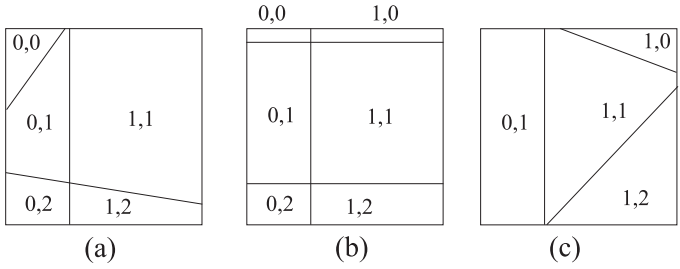} \caption{Three different
probability distributions for the outcomes of Alice and Bob
$(a_j=0,1; b_k=0,1,2)$, that satisfy the no-signaling condition.
The probability of each outcome is measured by the area of the
corresponding sector of the unit square. The marginal
distributions $P(a_j)$ and $P(b_k)$ are the same in the three
cases, while the correlations are completely different; (b)
describes the absence of correlation.} \label{carres}
\end{center}
\end{figure}

Now, in the case of two particles, a wide range of alternative
models can be made compatible with the no-signaling condition.
This can be easily seen by looking at a graphic representation. In
fig. \ref{carres} we represent three possible probability
distributions for the joint outcomes of Alice and Bob; a
probability is measured by the area of the corresponding sector of
the unit square. We see that very different situations can be met
while conserving the same marginal distributions, that is, without
violating the no-signaling condition. In particular, a model that
predicts a {\em complete loss of correlations} is possible: if
$P(a_j,b_k)=P^{QM}(a_j)P^{QM}(b_k)$ for all measurements, clearly
the marginal distribution on each side is independent on what is
done on the other side, and agrees with the prediction of QM.

Thus, as long as one considers only correlation between two
particles, the no-signaling condition puts practically no
constraint on the correlations. However, two-particle correlations
can be seen as marginal distributions of a three-particle
probability distribution. This will be discussed in detail in the
next sections, but we want to give the reader the feeling of what
happens. We have now a probability distribution $P(a_j,b_k,c_l)$.
The no-signaling condition forces $P(a_j)$, $P(b_k)$ and $P(c_l)$
to agree with the corresponding prediction of QM, just as above.
But in addition, the no-signaling condition may also force some
{\em correlations} to agree with those predicted by QM.
Specifically, in the next section we shall build, for two
alternative models, a situation in which both $P_{AC}(a_j,c_l)$
and $P_{BC}(b_k,c_l)$ must be identical to those predicted by QM,
while $P_{AB}(a_j,b_k)$ is free {\em a priori}. Consider then the
simple case of a dichotomic measurement on each of the three
particles, for which QM predicts perfectly correlated outcomes:
$P^{QM}(000)=P^{QM}(111)=\demi$. For such a measurement, obviously
no freedom is left on $P_{AB}(a_j,b_k)$ if all the other marginals
must agree with QM: no alternative model can predict anything else
than $P_{AB}(00)=P_{AB}(11)=\demi$, $P_{AB}(01)=P_{AB}(10)=0$.

\section{Predictions of two alternative models}
\label{models}

\subsection{Model 1: preferred frame and finite velocity}

The first model that we consider is a natural modification of
Bohmian mechanics. In Bohmian mechanics, the outcomes of a
measurement are determined by local parameters (the local settings
of the measurement and the quantum potential in which the particle
propagates), and by a superluminal hidden communication between
the correlated particles \cite{bohm}. The hidden communication
takes place in a preferred frame (PF), an assumption that can be
made for quantum phenomena without contradicting relativity in the
macroworld \cite{pframe}. If the speed of this hidden
communication in the PF is $v_{hc}=\infty$, the predictions of
Bohmian mechanics reproduce perfectly those of QM. However, in the
same line of thought one may suppose that $v_{hc}<\infty$. In this
case, the model gives some predictions that differ from the QM
ones. In fact, the hidden communication sent by the first particle
that is detected may not reach the second particle before its own
detection. In this case, non-local correlations should disappear.
This defines the model.

As we demonstrated in section 2, a complete loss of correlations
can be made consistent with the no-signaling condition for two
entangled particles, that is, for the original EPR-Bohm
experiment. Eberhard \cite{ebe} was the first to explore the
predictions of such a model in the case of three particles: he
noticed that signaling could be allowed, at least in the
particular version of the model that he built.

\begin{figure}
\begin{center}
\epsfxsize=10cm \epsfbox{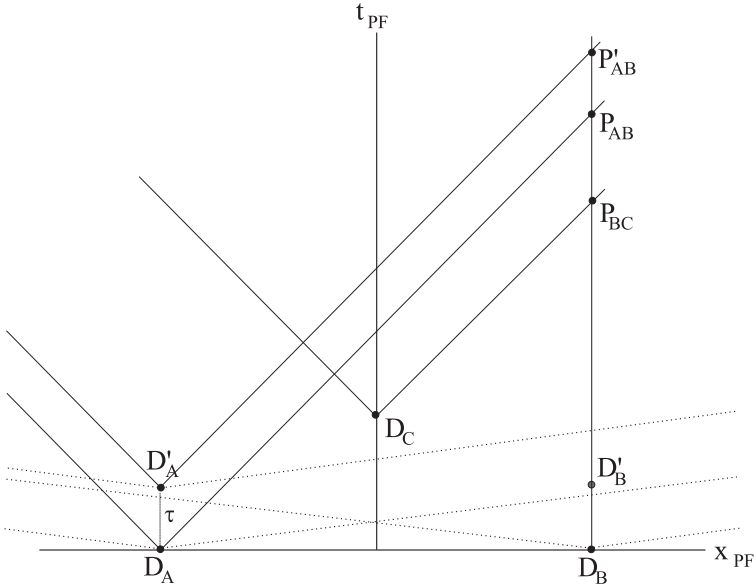} \caption{The events considered
for the argument in Model 1 (PF and $v_{hc}<\infty$). Events $D$
are detections. The black lines are light cones, the dotted lines
are the cones of the superluminal "hidden-communication". Bob can
calculate the correlations B-C in $P_{BC}$, before he can
calculate the correlation A-B: consequently the correlation B-C
cannot depend on whether Alice chose to put or not the time delay
$\tau$ on her detection. The argument can easily be made
symmetric.} \label{speed}
\end{center}
\end{figure}

We also consider a prediction of this model for an experiment
involving three particles. The three entangled particles are
labelled A, B and C, and we name Alice, Bob and Charlie the
experimenters that perform the measurements. The event "detection
of particle X" is written $D_X$. In the space-time coordinates of
the PF, we note \ba D_A\,=\,\big(-x,t_A\big)\,, &
\,D_B\,=\,\big(x,t_B\big)\,,&\,D_C\,=\,\big(0,t_C\big)\,. \ea If
the experimenters know which frame is the PF, they can arrange the
timings of the detection events in such a way that: (I) $D_A$ and
$D_B$ are simultaneous: $t_{A}=t_{B}=0$; (II) particle C is
detected later, and we suppose that the following condition
holds:\be \frac{3x}{v_{hc}}\,<\,t_{C}\,<\,\frac{x}{c}\,.
\label{tc}\ee The upper bound simply means that no classical
signal sent by A or B at $t=0$ might have reached C at $t_C$; the
lower bound is justified below. Condition (\ref{tc}) is not
contradictory as long as $v_{hc}>3c$, which cannot be taken as
absolutely granted since we don't know the PF. However, for two
reasonable candidates of PF (the laboratory frame, and the frame
of the cosmic background radiation), recent experiments provided a
lower bound for $v_{hc}$ which is several orders of magnitude
higher than $c$ \cite{sca}. So let's assume that $v_{hc}>3c$
\cite{referee}. Condition (\ref{tc}) implies that particle C has
received the hidden information from both A and B, since
$t_C>\frac{x}{v_{hc}}$, but no classical information. We claim
that the correlations A-C and B-C must be those predicted by QM,
in order to avoid signaling.

To prove this claim, we refer to fig. \ref{speed}, in which the
argument is sketched for the correlation B-C. $E(r_Br_C)$ can be
calculated by Bob after the space-time point
$P_{BC}=(x,t_C+\frac{x}{c})$. At that time, Bob cannot have
received any classical information from Alice, since an
information propagating from $D_A$ to $D_B$ at the speed of light
cannot arrive before $P_{AB}=(x,\frac{2x}{c})$. Therefore, the
correlation $E(r_B r_C)$ cannot change whatever Alice does,
otherwise we would allow signaling from Alice to the pair
Bob-Charlie.

Suppose now that Alice delays the detection of particle A by a
time $\tau=\frac{x}{v_{hc}}+\epsilon$, so that the event
"detection of A" is now $D_A'$. In this case, A receives the
hidden information from B, and C still receives from both --- this
is the justification of the lower bound in (\ref{tc}). So now we
have a normal time ordering $B\rightarrow A\rightarrow C$, thus QM
should apply. Consequently, the only way to avoid signaling is to
force correlation B-C to be the one predicted by QM, independently
of whether A has or has not received the hidden information from
B. A completely symmetrical argument, allowing Bob to delay his
detection to $D'_B$, leads to the conclusion that the correlation
A-C must also be the QM one.

Now, what about correlation A-B? Within the model, it depends on
the situation: (i) If the detections are simultaneous, i.e. in
cases $(D_A,D_B)$ or $(D'_A,D'_B)$, then A and B are either
uncorrelated (if we suppose that correlations arise only through
reception of the hidden communication), or at most correlated
through local variables; (ii) If one detection is delayed, i.e. in
cases $(D'_A,D_B)$ or $(D_A,D'_B)$, this correlation should agree
with QM.

In conclusion: under some circumstances this model predicts the
disappearance of non-local correlations between A and B, while A-C
and B-C are correlated according to QM.

\subsection{Model 2: Multisimultaneity}

This model was proposed some years ago by Suarez and one of us
\cite{sua}, and developed by Suarez in subsequent papers, where he
called it "Multisimultaneity" \cite{sua2}. Here there is not a
unique PF. Rather, there are several meaningful frames: for each
particle, the meaningful frame is the rest frame of the massive
{\em choice-device} that it meets. The choice-device may be the
detector --- in which case the "choice" is a form of "collapse"
--- or the beam-splitter, again in a Bohm-like interpretation where the
particle is really localized in a path, its "choices" being
determined by a pilot wave or quantum potential. Experiments have
been performed with detectors \cite{zbi} and with beam-splitters
in motion \cite{stefanov}, both vindicating QM. Our argument is
independent of what a choice-device is.

For two particles, Multisimultaneity works as follows: when a
particle meets a choice-device, it considers whether {\em in the
rest frame of this device} the other particle has already met its
own device \cite{note1}. Thus, according to the state of motion of
the choice-devices, we can arrange different timings: (I) {\em
Before-before timing}, in which each particle arrives to its own
choice-device, in the rest frame of this device, before the other
one. In this case, each particle chooses its outcomes taking into
account only local settings: two-particle correlations should
disappear completely; (II) {\em Before-after timing}, in which the
"before" particle chooses randomly its output, while the "after"
one takes into account the measurement that has been performed on
the first one. In the before-after timing, Multisimultaneity
reproduces the QM results; and this is the situation achieved in
all standard EPR experiments \cite{asp,tittel,weihs}, that
supported QM. (III) {\em After-after timing}, which is a rather
troubling feature of this kind of theories, and is not easy to
define while avoiding the causal loop \cite{sua2,zbi}.

\begin{figure}
\begin{center}
\epsfxsize=6cm \epsfbox{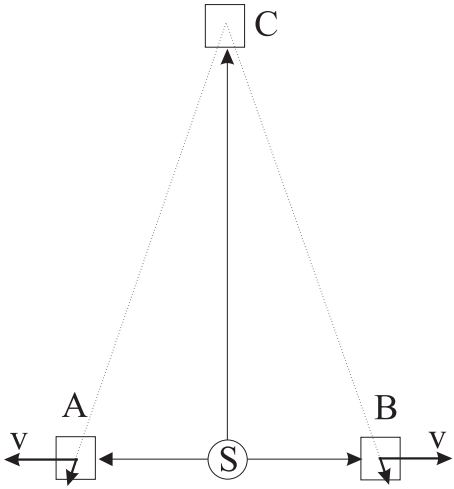} \caption{Experimental setup for
the argument in Model 2 (Multisimultaneity). The squares represent
the "choice devices". See text for details.} \label{multi}
\end{center}
\end{figure}

Again, let's turn to three particles. Consider the experimental
setup sketched in fig. \ref{multi}. Three particles issued from
the source S reach the respective choice-devices. The
choice-devices of A and B can be put into opposite motion with
speed $v$, high enough to achieve a before-before configuration
\cite{sua}. Therefore, when both devices are at a relative rest,
QM should apply; when they are in relative motion, particles A and
B have to choose in a before-before configuration, therefore their
non-local correlation should disappear. As for particle C: its
choice-device is motionless in the laboratory, and the projection
of the speed $v$ on the lines A-C and B-C (small black arrow in
the figure) is too small to change the timing. To be concrete, we
suppose that when C meets its choice-device, A and B have already
made their choice in the laboratory frame.

An argument analogous to the one discussed for Model 1 applies
here: correlations A-C and B-C must be in any case those predicted
by QM. In fact, suppose that Bob's device is at rest: Alice may
put her device into motion, and the pair Bob-Charlie should see no
difference. Symmetrically, if Alice's device is at rest, Bob
should be allowed to put his device into motion without signaling
to Alice-Charlie. Correlation A-B is the QM one when both Alice's
and Bob's devices are at rest in the laboratory frame; but when at
least one device moves, the non-local correlations should
disappear. Thus, Multisimultaneity also predicts that in some
cases A-B can be at most correlated through local variables, while
A-C and B-C are correlated according to QM.

\section{Hidden communication alone implies signaling}

In the context of alternative models invoking superluminal hidden
communication, the content of section \ref{twothree} can be
rephrased as follows. In the case of two particles, it is possible
to imagine a model in which the correlations are due only to the
hidden communication: if the communication is received, the
outcomes of the measurements becomes correlated; if the
communication is not received, the outcomes are uncorrelated. Here
we show that this is no longer true when we go to three particles.

To simplify the analysis, we assume that on each of the three
particles a local dichotomic measurement is performed. Alice's
possible results $r_A$ are labelled $\xi_A=\pm 1$, and similarly
for Bob's and Charlie's measurements. With this labelling, any
probability distribution can be written \ba p(r_A=\xi_A,
r_B=\xi_B,r_C=\xi_C)&=&\frac{1}{8}\,\Big[ 1+\xi_A E(r_A)+\xi_B
E(r_B)+\xi_C E(r_C)+\xi_A\xi_B E(r_Ar_B)+\nonumber\\
&&+\xi_A\xi_C E(r_Ar_C)+\xi_B\xi_C E(r_Br_C)+\xi_A\xi_B\xi_C
E(r_Ar_Br_C) \Big]\,, \label{probasgen}\ea where $E(.)$ is the
expectation value of the random variable. We have seen that both
Models 1 and 2 predict a configuration in which:

{\bf Condition 1:} $E(r_A)$, $E(r_B)$, $E(r_C)$, $E(r_Ar_C)$ and
$E(r_Br_C)$ must be those predicted by QM; this is the
no-signaling condition in the configurations that we have built.

Moreover, we are supposing that the correlations arise only
through reception of the hidden communication, Therefore we add to
Models 1 and 2 the following:

{\bf Condition 2:} $E(r_Ar_B)=E(r_A)E(r_B)$: complete loss of
correlation.

We provide now examples of quantum states and measurements for
which Conditions 1 and 2 lead to negative probabilities. Since we
consider dichotomic observables, the meaningful degrees of freedom
of each particle can be mapped onto a two-dimensional quantum
system (qubit).

{\em First example:} This is the example that we have already
discussed in a semi-qualitative way in section 2. Take the GHZ
state $\frac{1}{\sqrt{2}}(\ket{000}+\ket{111})$ \cite{ghz}, with
$\ket{0}$ and $\ket{1}$ the eigenstates of $\si_z$. If we measure
$\si_z$ on all particles, the quantum statistics give
$E^{QM}(r_A)=E^{QM}(r_B)=E^{QM}(r_C)=0$,
$E^{QM}(r_Ar_C)=E^{QM}(r_Br_C)=1$. Condition 2 gives consequently
$E(r_Ar_B)=0$. If we put this into (\ref{probasgen}), we find in
particular $p(++-)=\frac{1}{8}[-1-E(r_Ar_Br_C)]$,
$p(--+)=\frac{1}{8}[-1+E(r_Ar_Br_C)]$. These numbers cannot be
both non-negative: there is no three-particle probability
distribution that is compatible with both Conditions 1 and 2.
Actually for this particular case the only way to have
non-negative probabilities is to set $E(r_Ar_B)=1$ and
$E(r_Ar_Br_C)=0$, exactly the values predicted by QM. Thus this
example shows that all possible models that would predict a
reduction of visibility for all measurements, that is
$E(r_Ar_B)=V\,E^{QM}(r_Ar_B)$ with $0\leq V<1$, lead to signaling
--- we are not aware of any explicit model of this kind.

{\em Second example:} Take the state
$\ket{W}=\frac{1}{\sqrt{3}}(\ket{001}+\ket{010}+\ket{100})$
\cite{dur}. If we measure $\si_x$ on all particles, the quantum
statistics give $E^{QM}(r_A)=E^{QM}(r_B)=E^{QM}(r_C)=0$,
$E^{QM}(r_Ar_C)=E^{QM}(r_Br_C)=\frac{2}{3}$. Condition 2 gives
consequently $E(r_Ar_B)=0$. If we put this into (\ref{probasgen}),
we find in particular
$p(++-)=\frac{1}{8}[-\frac{1}{3}-E(r_Ar_Br_C)]$,
$p(--+)=\frac{1}{8}[-\frac{1}{3}+E(r_Ar_Br_C)]$. Again, these
numbers cannot be both non-negative. However here we could recover
non-negative probabilities by having $E(r_Ar_B)=\frac{1}{3}$,
which is lower than the QM value $E(r_Ar_B)=\frac{2}{3}$, but
still different from 0.

The appearance of negative probabilities means that the operation
"suppression of the correlation between some partners" is not a
completely positive (CP) map. It has been recently shown
\cite{simon} that if a system allows a description agreeing with
QM at a given time, the no-signaling condition forces the dynamics
to be described by a CP map.

\section{Consequences for alternative models}

In the introduction we explained that, in the classical world,
correlations seem to be always due either to preparation or to
communication. Bell's theorem ruled out the first of these origins
for EPR correlations. In this paper, we have ruled out also the
hypothesis of a hidden communication as unique cause of the
correlations, at least in the most natural alternative models.

One can still hope to develop an alternative model that introduces
{\em both} local hidden variables and hidden communication; we
call these {\em mixed models} \cite{durt}. In a mixed model,
correlations are established by hidden communication if this
communication is allowed to reach the other particles, or using a
pre-established information if for some reason the hidden
communication has not been received. We explain here the relation
between our approach and signaling in mixed models.

Consider again Example 1 of the previous section. For that
measurement, only the correlations predicted by QM are compatible
with the no-signaling condition. Unfortunately, no hidden
communication at all is needed to establish such correlations.
Even more generally: all the two-particle correlations predicted
by QM for a GHZ state can be described by classical mixtures,
since the two-particle partial trace on $\ket{GHZ}$ is
$\rho=\demi\ket{00}\bra{00}+\demi\ket{11}\bra{11}$. Consider then
the following mixed model: (i) the source produces an equal-weight
mixture of the possibilities $\ket{000}$ and $\ket{111}$. (ii) If
the hidden communication is allowed to reach the particles, the
remarkable three-particle GHZ correlations \cite{ghz} are
established. (iii) If the hidden communication does not reach the
particles, their correlations are determined by the preparation.
In this last case, the three-particle correlation will disagree
with the prediction of QM, but this is not in conflict with the
no-signaling condition as long as we have only three particles.

Of course, the question is, whether by considering other quantum
states one can show that also mixed models lead to signaling. To
do so, one should show that the no-signaling condition can force a
violation of a Bell's inequality. Let us sketch the idea of this
in the same configuration described above, in which A and B cannot
share the hidden communication and are therefore supposed to be
correlated only through preparation.

Alice makes on her side $N$ different measurements, and similarly,
Bob makes $M$ different measurements; for simplicity, we assume
that Charlie makes only one measurement. Thus the three partners
share $N\times M$ probability distributions
$P(a_j^{(n)},b_k^{(m)},c_l)$. The no-signaling constraint says
that $P(a_j^{(n)},c_l)$ and $P(b_k^{(m)},c_l)$ must be those
predicted by QM, whatever the measurements performed by Alice and
Bob; and this requirement puts some constraints on the marginals
$P(a_j^{(n)},b_k^{(m)})$; for instance, we have demonstrated above
that $P(a_j^{(n)},b_k^{(m)}) =P(a_j^{(n)})P(b_k^{(m)})$ is
impossible. Suppose now that there is a combination of these
two-particle probabilities that defines a Bell's inequality which
is violated by the QM probabilities $P^{QM}(a_j^{(n)},b_k^{(m)})$.
One can hope that the constraints put by the no-signaling
condition are strong enough to force the violation of the same
inequality, though possibly a weaker violation. If this is the
case, then no suitable preparation can lead to probabilities
$P(a_j^{(n)},b_k^{(m)})$ compatible with the no-signaling
constraints: mixed models would also imply signaling.

We have not found such a state and set of measurements. We note
that the study of the violation of Bell's inequalities by partial
states \cite{myprs}, as well as the study of Bell's inequalities
beyond the case of two qubits \cite{collins}, are presently open
fields of research. So it is likely that in the next years we
shall dispose of more adequate tools to tackle the robustness of
mixed models against signaling.

\section*{Acknowledgements}

We thank Ph. Eberhard, S. Popescu and A. Suarez for fruitful
discussion, and an anonymous referee for valuable comments. We
acknowledge partial financial support from the Swiss FNRS.


\begin{thebibliography}{10}
\bibitem{referee} A referee pointed out that one can find a
different ordering of the events such that the condition $v>3c$
can be eliminated, at least for the example of the GHZ state. The
situation is the following: particle $A$ leaves on the left,
particles $B_1$ and $B_2$ on the right. These two particles are
measured at nearby locations but simultaneously, so that they
cannot be correlated by the hidden communication. If both have
received the communication from particle $A$, they must become
correlated, because of the explanation at the end of section
\ref{twothree}. If they have not received communication from $A$
and if the correlation arise only through communication, then
$B_1$ and $B_2$ are uncorrelated. Note that this is a different
geometry from the one proposed in the main text. Here, the
alternatives are "no pair correlated" vs "pairs $A-B_1$ and
$A-B_2$ correlated" (and forcing a correlation on $B_1-B_2$). In
the main text, the alternatives are "pairs $A-C$ and $B-C$
correlated" vs. "all three pairs correlated".
\bibitem{note1} Note that Multisimultaneity predicts differences
with QM also in the case $v_{hc}=\infty$. The case of finite
$v_{hc}$ has not been studied within this approach.
\end{thebibliography}
\end{document}